\documentclass[preprint,showpacs,aps,prd,nofootinbib,letterpaper]{revtex4-1}
\usepackage{graphicx}
\usepackage{subfigure}
\usepackage{dcolumn}
\usepackage{amsmath}
\usepackage{amssymb}
\usepackage{times}
\usepackage{epsfig}
\usepackage{float}     
\usepackage{relsize}
\usepackage{float}
\RequirePackage{xspace}

\begin{document}
\title{The  $ J/\psi$ production in PbPb ultraperipheral collisions  at $\sqrt{s_{NN}}=2.76\mathrm{TeV}$}
\author{Ya-ping Xie}\email{xieyaping@impcas.ac.cn}
\affiliation{Institute of Modern Physics, Chinese Academy of
Sciences, Lanzhou 730000, China}
\affiliation{Department of Physics, Lanzhou University, Lanzhou 730000, Chnia}
\affiliation{Key Laboratory of Quark and Lepton Physics (MOE) and Institute
 of Particle Physics,
Central China Normal University, Wuhan 430079, China
}
\author{Xurong Chen}\email{xchen@impcas.ac.cn}
\affiliation{Institute of Modern Physics, Chinese Academy of
Sciences, Lanzhou 730000, China}
\begin{abstract}
We calculate the coherent and incoherent production of $J/\psi$ in PbPb ultraperipheral collisions. The production of $J/\psi$ in ultraperipheral collsions is product of photon flux distributions and cross section of photon-nucleus scatterings. The distributions of photon flux is computed in light-cone perturbation theory and the cross section of photon-nucleus scatterings is calculated in dipole model, we assume that the two gluons exchange contribution is the coherent cross section and the  large-$N_c$ contribution is the incoherent cross section in photon-nucleus scattering. The numerical result of the rapidity distributions of $J/\psi$ production in PbPb ultraperipheral collisions at $\sqrt{s_{NN}}=2.76$TeV are compared with the experimental data measured by the ALICE collaboration.   
\end{abstract}
\maketitle
 \section{Introduction}
The ultraperipheral collisions of relativistic heavy ions provide an opportunity to study the   nucleus-nucleus interaction at  high energies~\cite{Bertulani:2005ru,Baltz:2007kq}. The heavy quarkonia  can be produced in ultraperipheral relativistic heavy ion collisions. The $J/\psi$ photonproductions  in PbPb and p-Pb ultraperipheral collisions have  recently been measured  by the ALICE collaboration at CERN Large Hadron Collision(LHC) and the experimental data have been published in Refs.~\cite{Abelev:2012ba,Abbas:2013oua,TheALICE:2014dwa}. On the other hand, many theoretical groups have studied the production of heavy vector mesons in ultraperipheral collisions at high energies using different approaches~\cite{Frankfurt:2002sv,Klein:1999qj,Rebyakova:2011vf,Klein:2000dk,Klein:2002wm,Ducati:2013bya,Dominguez:2011cy,Adeluyi:2012ph,Adeluyi:2013tuu,Goncalves:2009kda,Goncalves:2011vf,Lappi:2013am,Guzey:2013qza}.\\
\indent In the nucleus-nucleus collisions, when the impact parameter of the two nucleus is larger than the sum of their radius, there is no hadronic interactions between the two nucleus. The nuclei can interact by  photons exchange. As photons can be emitted from nuclei at  high energies. This collisions are called ultraperipheral collisions. Two types of photon scattering can occur in the ultraperipheral collisions. The first process is photon-nucleus scattering. The second one is photon-photon scattering. We only consider the vector meson production in the photon-nucleus scattering in this work. \\
\indent The process of photon-nucleus scattering is well described in dipole mode~l\cite{Mueller:1993rr,Mueller:1994jq,Caldwell:2010zza} including the  wave functions of photon and  vector meson~\cite{Brodsky:1994kf,Munier:2001nr} in small-$x$ physics. In dipole model, the photon can fluctuate into a dipole of quark and antiquark, and the dipole scatters on the nucleons by gluons exchange. There are some models of parameterization for the cross section between the dipole and nucleons. For example GBW model~\cite{GolecBiernat:1998js,GolecBiernat:1999qd}, IIM model~\cite{Iancu:2003ge,Watt:2007nr} and IPsat model~\cite{Kowalski:2003hm,Kowalski:2008sa,Lappi:2010dd}. The wave function of photon can be calculated in light-cone perturbation theory, and the wave function of the vector meson can be parameterized in Gaus-LC and boosted-Gauss mode~l\cite{Kowalski:2006hc}.\\ 
\indent In  the process of photon-nucleus scattering, the nucleus can remain intact or break up. If the nucleus  remains intact, it is  coherent process. If the nucleus breaks up, it is  incoherent process. The authors  of Refs.~\cite{Lappi:2010dd,Lappi:2013am} presented the coherent and incoherent cross section .But, the ratio of incoherent to  coherent of Ref.~\cite{Lappi:2013am} is lower than the experimental ratio. In this work, we  distinguish coherent and incoherent cross section in a new mechanism, and we think the contribution of our calculation also contribute the ratio of the incoherent to the coherent production.\\
\indent This paper is organized as follows: In Sec II, we present the calculation of distributions of the equivalent photon flux. In Sec III, the coherent and incoherent cross section are considered from color-dynamics, the numerical results   are presented in Sec IV. The conclusions of this paper is in Sec IV.
\section{The distributions of equivalent photon flux }
\begin{figure}[!h]
\begin{center}
\includegraphics[width=3in]{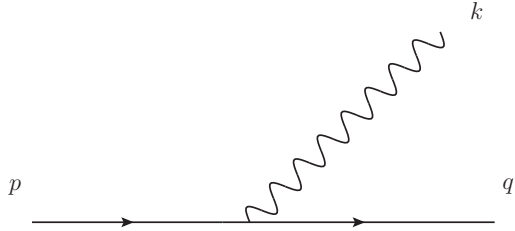}\hspace{3.0cm}
\caption{The diagram of the emission of photon from a nucleus.}\label{fig1}
\end{center}
\end{figure}

\indent  In Jackson's textbook~\cite{Jackon}, the  distributions of equivalent photon flux  were calculated in  classical electrodynamics, we shall calculate the distributions of equivalent photon flux in light cone perturbation theory in this work.  We start with a photon emitted  from a nucleus with $Z$ electric charges, the  emission of photon from  a nucleus is illustrated in Fig.~\ref{fig1}.  We can write down their momentums in light-cone conventions, they are
\begin{eqnarray}
p=(p^+,\frac{p_\perp^2+m^2}{2p^+}, p_\perp), 
q=((1-\chi )p^+, \frac{(p_\perp -k_\perp)^2+m^2}{2(1-\chi )p^+}, p_\perp -k_\perp), 
k=(\chi p^+, \frac{k_\perp^2}{2\chi p^+},k_\perp).
\end{eqnarray}
The variable $p$ is the momentum of the initial charged nucleus,  $p^+=\frac{p_0+p_3}{\sqrt{2}}=\sqrt{2}E$, where $E$ is the energy of the nucleus, and $k$ is the momentum of the photon, $k^+=
\frac{k_0+k_3}{\sqrt{2}}=\sqrt{2}\omega$ , where $\omega $ is the energy of the photon,  and $q$ is the momentum of the final charged nucleus, $m$ is the mass of the nucleus, $\chi $ is the  momentum fraction of the initial nucleus carried by the photon. 
The  cross section of the nucleus-nucleus collisions by photon exchange can be written as
\begin{eqnarray}
\sigma(AA)=\sum_{\lambda\alpha\beta}\int \frac{d^3k}{2(2\pi)^3}|\psi^\lambda_{\alpha\beta }(p,k )|^2
\sigma(\gamma A),
\label{ssigma}
\end{eqnarray}
where $\sigma(\gamma A)$ is the cross section of photon-nucleus scattering.  The $\psi^\lambda_{\alpha\beta }(p,k )$ is the splitting wave function of the nucleus and  photon, the splitting wave function in momentum space can be  written as
\begin{eqnarray}
\psi_{\alpha\beta }^{\lambda}(p,k)=\frac{Ze}{\sqrt{8p^+k^+(p-k)^+}}\frac{\bar{u}_\beta(p-k)\gamma_\mu\cdot\epsilon^\mu_{\lambda}(k)
u_\alpha(p)}{(p-k)^-+k^--p^-},
\end{eqnarray}
where $\lambda$ denotes indice of polarization vector, $\alpha $ and $\beta$ denote the helicities  of incoming and outgoing nucleus, and $\epsilon^\mu_{\lambda}(k)$ is the polarization vector of the photon, where $\epsilon_\lambda^\mu(k)=(0,\frac{k_\perp\cdot \epsilon_\perp^\lambda}{k^+},\epsilon^\lambda_\perp)$,  with $\epsilon^{\lambda=1,2}_\perp=\frac{1}{\sqrt{2}}(\mp1,i)$.  With the help of the light-cone theory, we can write the splitting wave function in momentum space as~\cite{Marquet:2007vb}
\begin{eqnarray}
\psi^\lambda_{\alpha\beta }(p,k )=&&\frac{Ze}{\sqrt{k^+}}
\frac{1}{(k_\perp-\chi p_\perp)^2+\chi^2m^2}\notag\\&&\times\begin{cases}
\sqrt{2} (k_\perp-\chi p_\perp)\cdot\epsilon_\perp^\lambda\left[\delta_{\alpha-}\delta_{\beta-}+(1-\chi )\delta_{\alpha+}\delta_{\beta+}\right]+m\chi ^2\delta_{\alpha+}\delta_{\beta-}, 
& \lambda=1, 
\\
\sqrt{2} (k_\perp-\chi p_\perp)\cdot\epsilon_\perp^\lambda\left[\delta_{\alpha+}\delta_{\beta+}+(1-\chi )\delta_{\alpha-}\delta_{\beta-}\right]-m\chi ^2\delta_{\alpha-}\delta_{\beta+}, 
& \lambda=2,
\end{cases}
\end{eqnarray}
  The splitting wave function can be written in the coordinate space, which is fourier transform of the wave function in momentum space, 
\begin{eqnarray}
\psi^\lambda_{\alpha\beta }(p,k^+;\mathbf{x} )=\int d^2k_\perp e^{k_\perp\cdot \mathbf{x}}\psi^\lambda_{\alpha\beta }(p,k^+).
\label{fours2}
\end{eqnarray}
The splitting wave function of coordinate space reads
\begin{eqnarray}
&&\psi^\lambda_{\alpha\beta }(p,k^+;\mathbf{x} )=\frac{2Ze\pi m}{\sqrt{k^+}}
e^{i\chi p_\perp\cdot \mathbf{x} }\notag\\&&\times\begin{cases}
i\chi \sqrt{2}K_1(\chi m\vert\mathbf{x}\vert ) \mathbf{x} \cdot\epsilon_\perp^\lambda/\vert\mathbf{x}\vert \left[\delta_{\alpha-}\delta_{\beta-}+(1-\chi )\delta_{\alpha+}\delta_{\beta+}\right]+\chi ^2K_0(\chi m\vert\mathbf{x}\vert)\delta_{\alpha+}\delta_{\beta-}, 
& \lambda=1, 
\\
i\chi \sqrt{2}K_1(\chi m\vert\mathbf{x}\vert) \mathbf{x} \cdot\epsilon_\perp^\lambda/| \mathbf{x}|\left[\delta_{\alpha+}\delta_{\beta+}+(1-\chi )\delta_{\alpha-}\delta_{\beta-}\right]-\chi ^2K_0(\chi m| \mathbf{x}| )\delta_{\alpha-}\delta_{\beta+}, 
& \lambda=2,
\end{cases}
\label{wavecs}
\end{eqnarray}
where  $K_0(x)$ and $K_1(x)$ are modified Bessel functions.

we can write down the splitting wave function in momentum space as fourier transform of splitting wave function in coordinate space,
\begin{eqnarray}
\psi^\lambda_{\alpha\beta }(p,k^+)=\int\frac{d^2\mathbf{x}}{(2\pi)^2}e^{-ik_\perp\cdot \mathbf{x}}\psi^\lambda_{\alpha\beta }(p,k^+;\mathbf{x} ).
\label{fs}
\end{eqnarray}
Substituting Eqs.~(\ref{wavecs}) and (\ref{fs}) into Eq.~(\ref{ssigma}), we get
\begin{eqnarray}
\sigma(AA\to J/\psi AA)=&&\sum_{\lambda\alpha\beta}\int \frac{d^3k}{2(2\pi)^3}|\psi^{\lambda}_{\alpha\beta}(p,k)|^2
\sigma(\gamma A\to J/\psi A)\notag\\
=&&\sum_{\lambda\alpha\beta}\int \frac{d^2k_\perp dk^+}{16\pi^3}
\int \frac{d^2\mathbf{x}}{(2\pi)^2} e^{-ik_\perp\cdot \mathbf{x}}\psi^\lambda_{\alpha\beta }(p,k^+;\mathbf{x} )\int \frac{d^2\mathbf{x^\prime} }{(2\pi)^2} e^{ik_\perp\cdot \mathbf{x}^\prime}\psi^\lambda_{\alpha\beta }(p,k^+;\mathbf{x^\prime} )\sigma(\gamma A\to J/\psi A)\notag\\ =&&\int\frac{ d^2\mathbf{x}dk^+}{k^+} \frac{Z^2\alpha }{2\pi^2}
\frac{(\chi m\mathbf{x})^2}{\mathbf{x}^2}\big\{K^2_1(\chi m\vert\mathbf{x}\vert)[1+(1-\chi )^2]+\chi^2K_0^2(\chi m \vert\mathbf{x}\vert)\big\}\sigma(\gamma A\to J/\psi A),
\end{eqnarray}
  with $ dk^+/k^+=d\omega/\omega$,  we get the cross section of the nucleus-nucleus scattering by photon exchange 
\begin{eqnarray}
\sigma(AA\to J/\psi AA)=\int\frac{d\omega d^2\mathbf{x} }{\omega }
N(\vert\mathbf{x}\vert,\omega)\sigma(\gamma A),
\label{dAA}
\end{eqnarray}
where  the distribution  $N(\vert\mathbf{x}\vert,\omega)$ is 
\begin{eqnarray} 
N(\vert\mathbf{x}\vert,\omega)=\frac{Z^2\alpha }{2\pi^2}
\frac{(\chi m\mathbf{x})^2}{\mathbf{x}^2}\big\{K^2_1(\chi m\vert\mathbf{x}\vert)[1+(1-\chi )^2]+\chi^2K_0^2(\chi m \vert\mathbf{x}\vert)\big\}.\label{dndkdr}
\end{eqnarray}
\indent From the classical electrodynamics~\cite{Jackon}, it is supposed that a charge $ze$
passes the origin at speed $v$ and impact parameter $x$, the electric field are
$E_T(t)=\frac{zex}{(x^2+\gamma_L^2v^2t^2)^{1/2}}$ and $E_L(t)=-\frac{ze\gamma_L vt}{(x^2+\gamma_L^2v^2t^2)^{1/2}}$, the fourier transform for frequency spectra is
 $E(\omega)=\frac{1}{\sqrt{2\pi}}\int_{-\infty}^{\infty}E(t)e^{i\omega t}dt$,
 the frequency spectra of electric field are $E_T(\omega)=\frac{ze}{xv}(\frac{2}{\pi})^{1/2}\omega x/\gamma_Lv K_1(\omega x/\gamma_Lv)$ and $E_L(\omega)=-i\frac{ze}{\gamma_Lxv}(\frac{2}{\pi})^{1/2}\omega x/\gamma_Lv K_0(\omega x/\gamma_Lv)$, where  $v\approx c\approx 1$ in ultraperipheral collisions, the equivalent  photon flux per unit area is calculated as
 \begin{eqnarray}
 N(\vert\mathbf{x}\vert,\omega)&=&\frac{1}{8\pi^2}[E_T^2(\omega)+E_L^2(\omega)]\notag\\
& =&\frac{Z^2\alpha}{\pi^2}
 \frac{(\omega |\mathbf{x}|/
 \gamma_L)^2}{\mathbf{x}^2}\big[K^2_1(\omega |\mathbf{x}|/
 \gamma_L)+\frac{1}{\gamma_L^2}K_0^2(\omega |\mathbf{x}|/
 \gamma_L)\big].\label{cdndkdr}
 \end{eqnarray}
  The lorentz boost factor $\gamma_L$  in the collision reads $\gamma_L=E/m$,  with $\chi =\omega/E$, we can write  $\chi m|\mathbf{x}|=\omega\vert\mathbf{x}\vert/\gamma_L$. 
  In the ultraperipheral collisions,  it is easy to get $\chi \ll1$, and $\gamma_L\gg 1$, we can neglect $\chi ^2K^2_0(\chi m\vert\mathbf{x}\vert)$  and  $\frac{1}{\gamma_L^2}K_0^2(\omega |\mathbf{x}|/
   \gamma_L)$ in  Eq.~(\ref{dndkdr})
    and Eq.~(\ref{cdndkdr}). The dominant of distribution of $N(|\mathbf{x}|,\omega)$ is 
    \begin{equation}
 N(\vert\mathbf{x}\vert,\omega )=\frac{Z^2\alpha}{\pi^2}
 \frac{(\omega |\mathbf{x}|/
 \gamma_L)^2}{\mathbf{x}^2}K^2_1(\omega |\mathbf{x}|/
 \gamma_L).
    \end{equation}
    As to get the  usable photon flux $n(\omega)$,  we  integrate  $N(x, \omega) $ over the $\mathbf{x}$ and its angle $\theta$ as Ref.~\cite{Cahn:1990jk}, in the ultraperipheral collisions,  with $\vert\mathbf{x}\vert>2R_A$, $\vert\mathbf{x}\vert_{min}=2R_A$, and $R_A$ is the radius of the nucleus, we get  
  \begin{eqnarray}
n(\omega)&=&\int_0^{2\pi} d\theta\int_{2R_A}^{\infty}\vert\mathbf{x}\vert d\vert\mathbf{x}\vert N(x, \omega)\notag\\
&=&\int_{2R_A}^{\infty}\vert\mathbf{x}\vert d\vert\mathbf{x}\vert\frac{2Z^2\alpha}{\pi}
 \frac{(\omega |\mathbf{x}|/
 \gamma_L)^2}{\mathbf{x}^2}K^2_1(\omega |\mathbf{x}|/
 \gamma_L),
\end{eqnarray}
the integration result is    
\begin{eqnarray}
n(\omega)=\frac{2 Z^2\alpha}{\pi}\big[\xi K_1(\xi)
K_0(\xi)-\frac{\xi^2}{2}[K_1^2(\xi)-K_0^2(\xi)]\big],
\end{eqnarray}
 where   $\xi=2\omega R_A/\gamma_L$, thus, the Eq.~(\ref{dAA}) can be written as 
 \begin{equation}
\sigma(AA\to J/\psi AA)=\int d\omega \frac{n(\omega)}{\omega }
\sigma(\gamma A).
 \end{equation}
\indent With $\omega=\frac{M_V}{2}\exp(y)$, we get $d\omega/\omega=dy$, and $y$ is the rapidity of the vector meson, with mass  $M_V$. Finally, we get the rapidity distributions of the vector  meson~\cite{Lappi:2013am}
\begin{eqnarray}
\frac{d\sigma^{A_1A_2}}{dy}=n^{A_1}(y)\sigma^{\gamma A_2}(y)+n^{A_2}(-y)\sigma^{\gamma A_1}(-y),
\label{dndk}
\end{eqnarray}
where  $\sigma^{\gamma A}(y)$ is the cross section of the photon-nuclues scattering, and $n(y)$ is  the distribution of the photon flux, we shall calculate the cross section of photon-nucleus scattering at the next section.
\section{The cross section of coherent and incoherent processes}
\subsection{dipole model }
\begin{figure}[h]
\begin{center}
\includegraphics[width=3in]{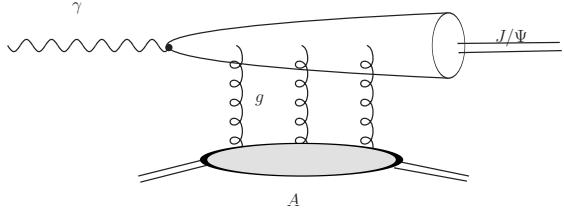}\hspace{3.0cm}
\caption{The diagram of $\gamma +A\to J/\psi+ A$, the photon breaks into dipole and the dipole scatters on the nucleons, and the dipole becomes vector meson.}\label{fig2}
\end{center}
\end{figure}

\indent After we calculate the rapidity distributions of the vector mesons in the ultraperipheral collisions, we only calculate the photon-nucleus cross section $\sigma^{\gamma A}(k)$ of Eq.~(\ref{dndk}). The process of photon-nucleus scattering is illustrated in Fig.~\ref{fig2}. In small-$x$ physics, the dipole model describes the scattering of the photon-nucleus successfully. It is shown in Fig.~\ref{fig2} that the process of $\gamma A\to V A$ can be viewed as three steps. The photon breaks up into a pair of quark and antiquark at first step, the quark and antiquark are called dipole, the dipole scatters on  the nucleons of the nuclei by gluons exchange at the second step, finally, the dipole becomes a vector meson at  third step. \\
\indent  The  differential cross section of vector meson in the photon-nucleus scattering  can be written as~\cite{Mueller:1994jq,Kowalski:2003hm,Kowalski:2006hc}
\begin{eqnarray}
\frac{d\sigma^{\gamma A\to VA}}{dt}=\frac{R_g^2(1+\beta^2)}{16\pi}
\left|\mathcal{A}^{\gamma A\to VA}_{T,L}(x_A, Q^2,\Delta) \right|^2,\label{dsigma}
\end{eqnarray}
where  $T$ and $L$ denote the transverse and longitudinal amplitudes. The factor
$\beta $ is the ratio of the real part to the imagigary part of amplitude. It reads
\begin{equation}
\beta=\tan(\frac{\pi}{2}\lambda),
\end{equation}
where $\lambda$ is calculated as
\begin{equation}
\lambda=\frac{\partial \ln (\mathrm{Im}\mathcal{A}(s))}{\partial \ln s}.
\end{equation}
The factor $R_g^2$ reflects  the skewdness, it gives 
\begin{equation}
R_g=\frac{2^{2\lambda+3}}{\sqrt{\pi}}\frac{\Gamma(\lambda+5/2)}{\Gamma(\lambda+3)}.
\end{equation}
The amplitude of $\mathcal{A}^{\gamma A\to VA}_{T,L}(x_A, Q^2,\Delta) $ contains  three parts, the light cone wave function of photon fluctuating into   $q\bar{q}$ dipole, the differential cross section of the dipole scatter on the nucleons, and the wave function of dipole recombining a vector meson. The amplitude reads 
\begin{equation}
\mathcal{A}^{\gamma A\to VA}_{T,L}(x_A, Q^2,\Delta)=i\int d^2r\int_0^1\frac{dz}{4\pi}
\int d^2b(\Psi_V^*\Psi_{\gamma})_{T,L}(r,z)e^{-ib\cdot \Delta }\frac{d\sigma_{q\bar{q}}}{d^2b}\label{amp},
\end{equation}
where  $t=-\Delta^2$,  $\Delta $ is the transfer momentum between the dipole and nucleons.   Integrating  over $t$, we can get the cross section $\sigma^{\gamma A}(k)$. The   $x_A=M_V\exp(-y)/\sqrt{s_{NN}}$ is Bjorken variable, and  $-Q^2$ is the virtuality of the photon, $b$ is the impact parameter between the dipole and the nucleons, $r$ is the size of the dipole, and $z$ is the  momentum  fraction of the photon carried by the quark or antiquark. The $(\Psi_V^*\Psi_{\gamma})_{T,L}(r,z)$ is the overlap of the functions of vector meson and the photon, the wave function of photon can be computed in light cone perturbation theory, the scalar function of the vector meson in this work is Gaus-LC  model which can be found in Ref.~\cite{Kowalski:2006hc}. In this work, we only consider the transverse amplitude, the transverse overlap reads
\begin{eqnarray}
(\Psi_V^*\Psi_{\gamma})_T(r,z)=e_fe\frac{N_c}{\pi z(1-z)}\lbrace  m_f^2
K_0(\epsilon r)\phi_T(r,z)-(z^2+(1-z)^2)\epsilon K_1(\epsilon r)\partial_r
\phi_T(r,z)\rbrace ,
\end{eqnarray}
where $m_f$ is the mass of charm quark, $e_f$ is the electric charge of the charm quark, and $\epsilon=\sqrt{z(1-z)Q^2+m_f^2}$, $N_c$ is the number of the colors. The scalar function  $\phi_T(r,z)$  of Gaus-LC model reads 
\begin{eqnarray}
\phi_T(r,z)=N_T(z(1-z))^2\exp(-\frac{r^2}{2R_T^2}).
\end{eqnarray}
The scalar function of Boosted Gaussian model is 
\begin{eqnarray}
\phi_T(z,r)=N_Tz(1-z)\exp\big(-\frac{m_f^2\mathcal{R}^2}{8z(1-z)}-
\frac{2z(1-z)r^2}{\mathcal{R}^2}+\frac{m_f^2\mathcal{R}^2}{2}\big).
\end{eqnarray}
The parameters of $N_T$, $R_T^2$ and $\mathcal{R}^2$ we used are the same as Ref.~\cite{Kowalski:2006hc}.\\
\subsection{coherent  cross section}
\begin{figure}[H]
\begin{center}
\includegraphics[width=4in]{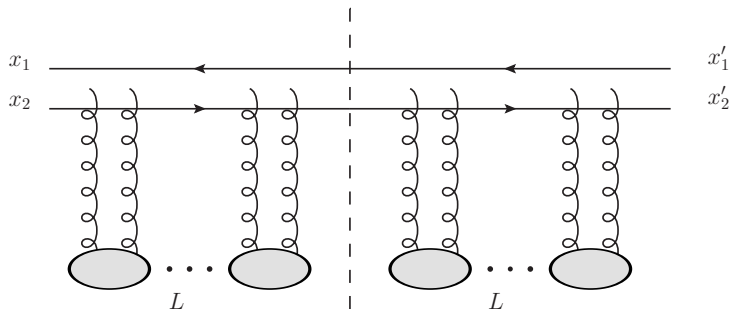}\hspace{3.0cm}
\caption{The diagram  of the scattering between the dipole and  nucleons exchanging  two-gluons including amplitude and conjugate amplitude, $L$ is the length that the dipole penetrates through the nucleus,  $x_1$ and $x_2$ are the coordinates of the quark and antiquark,  $x_1^\prime$ and $x_2^\prime$  are the conjugate coordinates.}\label{fig3}
\end{center}
\end{figure}

\indent  In the photon-nucleus scattering, the nucleus can remain intact or break up. If  the nucleus remains intact, it is coherent process. If the nucleus breaks up, it is incoherent process. We  consider the coherent and incoherent contribution to the vector meson production using dipole model. We assume that the nucleus is made of $A\gg 1$ nucleons, and the dipole scatters off the nucleons by gluons exchange. The interaction is assumed perturbative, therefore, the two gluons exchange contribution is the leading order contribution. The cross section between dipole and nucleons is illustrated in Fig.~\ref{fig3}. It is shown in Fig.~\ref{fig3} that the dipole scatters off the nucleons by two-gluon exchange. From QCD color dynamics, the color of two gluons can be color-singlet. The nucleus can remain color-singlet in the two-gluon exchange scattering. If the transfer momentum  satisfies $\Delta <1/R_A$, the nucleus can remain intact. Thus, we can treat the two gluons exchange contribution as coherent cross section.  \\
In the previous years, Golec-Biernat and W\"usthoff  proposed a model which described 
the dipole cross section successfully. The total dipole cross section was calculated
in the GBW model as~\cite{GolecBiernat:1998js,GolecBiernat:1999qd} 
\begin{equation}
\sigma_{q\bar{q}}(x,r)=\sigma_0(1-e^{-r^2Q_S^2(x)/4}),
\end{equation}
with $Q^2_S(x)=Q_{S0}^2(x_0/x)^\lambda$, $Q_{s0}^2=1\mathrm{GeV}^2$, the parameters $\sigma_0, x_0, \lambda $ are presented in Ref.~\cite{GolecBiernat:1998js}.
In this work, we use the GBW model as impact parameter dependent~\cite{Mueller:2002pi}.\\
The differential cross section reads
\begin{equation}
\frac{d\sigma_{q\bar{q}}}{d^2b}=2(1-S_{x_A}(x_1,x_2,b)),
\end{equation}
where  $S_{x_A}(x_1,x_2,b)$ is the element of the $S-$matrix, we assume it is real.
 The $x_1$, $x_2$ are coordinates of the quark and antiquark in the coordinate space, where  $r=x_1-x_2$. The  $S_{x_A}(x_1, x_2,b)$ is written in Ref.~\cite{Mueller:2002pi} 
\begin{equation}
S_{x_A}(x_1, x_2,b)=e^{-Q^2_S(x_A,b)(x_1-x_2)^2/4},
\end{equation}
where we use GBW model for the dipole amplitude, the  saturated momentum of proton is $Q_S^2(x_p)=Q_{S0}^2(x_0/x_p)^\lambda$, and
$Q_S^2(x_A)=A^{1/3}c(b)Q_{S0}^2(x_0/x_A)^\lambda$ for A nucleons~\cite{Stasto:2012ru}, if we consider the influences of impact parameter. It is modified as 
\begin{equation}
Q_S^2(x_A,b)=A^{1/3}Q_{S0}^2c(b)\sqrt{1-b^2/R_A^2}(x_0/x_A)^\lambda,
\end{equation}
where $c(b)$ is a parameter. Therefore,  we can write  $S_{x_A}(x_1,x_2,b)$ as  
\begin{eqnarray}
S_{x_A}(x_1,x_2, b)=\exp(\frac{-Q^2_{S0}A^{1/3}c(b)\sqrt{1-b^2/R_A^2}(x_0/x_A)^\lambda (x_1-x_2)^2}{4}).\label{sbs}
\end{eqnarray}
where $c(b)$, $\lambda$ and $x_0$  are parameters to be fit from $F2$. They are presented in next section. Then, we can write down the coherent differential cross section
\begin{eqnarray}
\frac{d\sigma^{\gamma A\to VA}}{dt}=\frac{R_g^2(1+\beta^2)}{16\pi}
\left|\int d^2r\int_0^1\frac{dz}{4\pi}
\int d^2b(\Psi_V^*\Psi_{\gamma})_{T} (r,z)e^{-ib\cdot \Delta }2(1-S_{x_A}(x_1,x_2,b))\right|^2,
\label{dsigma}\notag\\
\end{eqnarray}
where the  $S_{x_A}(x_1,x_2,b)$ is defined in  Eq.~(\ref{sbs}) and we only consider the real photon contribution in the ultraperipheral collisions.
\subsection{incoherent cross section}
\begin{figure}[H]
\begin{center}
\includegraphics[width=4in]{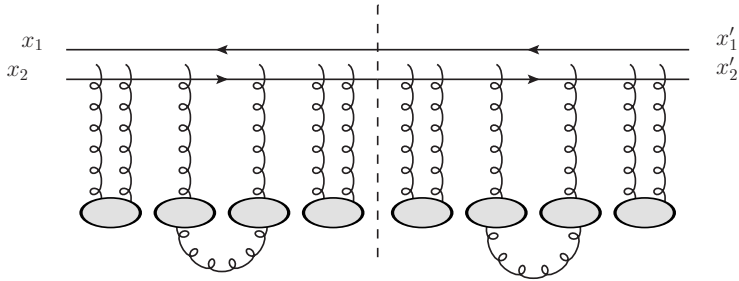}\hspace{3.0cm}
\caption{The diagram  of the scattering between the dipole and  nucleons exchanging  two-gluons including amplitude and conjugate amplitude, $L$ is the length that the dipole penetrates through the nucleus,  $x_1$ and $x_2$ are the coordinates of the quark and antiquark. $x_1^\prime$ and $x_2^\prime$  are the conjugate coordinates.}\label{fig5}
\end{center}
\end{figure}

\indent  Now, we consider double single gluon exchange cross section of the photon-nucleus scattering. The process with two single-gluon exchange is shown in Fig.~\ref{fig5}.  We can see that there are more two single gluon exchange  in Fig.~\ref{fig5}. 
The $\zeta$ or $\eta $ are the ratios between length from initial scattering to first or second single gluon exchange scattering and the total length $L$.  As to ensure that the final state vector meson is color-singlet, one single gluon exchange is impossible.  Two single-gluon exchange  
 can ensure  that the final state vector meson and the nucleus are color-singlet, but the nucleons are not all color-singlet. In the high energy limits, the size of the nucleus is $r\approx \frac{m_pR_A}{p^+}$, the parton's interaction radius is $r^\prime \approx 
 \frac{1}{xp^+}$,  in the ultraperipheral collision $x\ll \frac{1}{m_pR_A}$. The nucleons can exchange gluons easily after they emitted one gluon, the process is illustrated in the Fig.~\ref{fig5},  the two nucleons exchange a gluon each other.  The nucleons can ensure color singlet in the preocess, we think  the  contribution of the process depicted in Fig.~\ref{fig5} is part of the contribution of incoherent cross section in high energy limits.\\
The formulas of the differential cross section  with double single gluon exchange can be written as~\cite{Dominguez:2011wm}
	\begin{eqnarray}
	\frac{d\sigma^{\gamma A\to VX}}{dt}=&&\frac{R_g^2(1+\beta^2)}{4\pi}\int d^2r\int_0^1\frac{z}{4\pi}\int d^2b\int d^2r^\prime
			\int_0^1\frac{z^\prime}{4\pi}\int d^2b^\prime(\Psi_V^*\Psi_\gamma)_{T}(z,r)(\Psi_V^*\Psi_\gamma)_{T}(z^\prime ,r^\prime )\notag\\ &&\times e^{-i(b-b^\prime )\cdot \Delta} 
	\big(1-S_{x_A}(x_1,x_2,b)-S_{x_A}(x^\prime_2,x^\prime_1,b^\prime)+\langle S_{x_A}(x_1,x_2,b)S_{x_A}(x^\prime_2,x^\prime_1,b^\prime)\rangle\big),	\notag\\
	\label{co_cross}
	\end{eqnarray}
where  $\langle S_{x_A}(x_1,x_2,b)S_{x_A}(x^\prime_2,x^\prime_1,b^\prime)\rangle$ is  the dipole-dipole correlator~\cite{Blaizot:2004wv,Dominguez:2008aa}. It is also presented in Appendix of ~\cite{Dominguez:2011wm}, which can be factorized into the product of $ S_{x_A}(x_1,x_2,b)S_{x_A}(x^\prime_2,x^\prime_1,b^\prime)$ in MV model~\cite{McLerran:1993ni,McLerran:1993ka,McLerran:1994vd}. The detail calculation can be found in Refs.~\cite{Blaizot:2004wv,Dominguez:2008aa}.  We can write the dipole-dipole correlator as double integration of $\zeta$ and $\eta$ as follows:
\begin{eqnarray}
\langle S_{x_A}(x_1,x_2,b)S_{x_A}(x^\prime_2,x^\prime_1,b^\prime)\rangle=&&S_{x_A}(x_1,x_2,b)S_{x_A}(x^\prime_2,x^\prime_1,b^\prime)+S_{x_A}(x_1,x_2,b)S_{x_A}(x^\prime_2,x^\prime_1,b^\prime)\nonumber\\&&\times\frac{1}{N^2_c}[\frac{\mu^2N_c}{2}F(x_1,x_2;x^\prime _2,x^\prime_1)]^2 \int^1_0d\eta\int^\eta_0d\zeta e^{-\frac{\zeta\mu^2N_c}{2}F(x_1,x^\prime_2;x_2,x^\prime_1)},\notag\\
\label{dipoledipole}
\end{eqnarray}
where the function $F(x,y;u,v)$ is defined in Ref.~\cite{Dominguez:2008aa}. It reads
	\begin{equation} 
	\mu^2 F(x,y;u,v)\equiv\frac{Q_S^2}{2C_F}(x-y)\cdot(u-v).
	\end{equation}
The relationships of $x_1$ ,$x_2$ and $r$ ,$b$ are easy to get.  They can be written as 
	\begin{equation}
	\begin{split}
 	x_1=b+r/2,\quad
	x_2=b-r/2,\\
	x_1^\prime=b^\prime +r^\prime/2,\quad
	x_2^\prime=b^\prime -r^\prime/2.
			\end{split}
	\end{equation}
	We can get the two functions  as
	\begin{equation}
	\mu^2F(x_1,x_2;x_2^\prime ,x^\prime_1)=-\frac{Q_S^2}{2C_F}r\cdot r^\prime,
	\end{equation}
	and 
	\begin{equation}
	\mu^2F(x_1,x_2^\prime;x_2 ,x^\prime_1)=-\frac{Q_S^2}{8C_F}[(r+r^\prime )^2-4(b-b^\prime )^2].
	\end{equation}
Therefore, we can calculate the second line  of Eq.~(\ref{dipoledipole}) as 
	\begin{eqnarray}
	&&\frac{1}{N^2_c}[\frac{\mu^2N_c}{2}F(x_1,x_2;x^\prime _2,x^\prime_1)]^2 \int^1_0d\eta\int^\eta_0d\zeta e^{-\zeta\frac{\mu^2N_c}{2}F(x_1,x^\prime_2;x_2,x^\prime_1)}
\notag	\\ &&=
	\frac{16(r\cdot r^\prime )^2}{N_c^2((r+r^\prime )^2-4(b-b^\prime )^2)}[e^{\frac{Q_S^2((r+r^\prime )^2-4(b-b^\prime )^2)}{8}}-\frac{Q_S^2((r+r^\prime )^2-4(b-b^\prime)^2)}{8}-1].
	\label{ddc}
	\end{eqnarray}
	If  $b-b^\prime=0$, we get the same result as Eq.~(49) of Ref.~\cite{Dominguez:2008aa}.
Then, the  differential cross including large-$N_c$ contribution can be written as
	\begin{eqnarray}
	\frac{d\sigma^{\gamma A\to VX}}{dt}=&&\frac{R_g^2(1+\beta^2)}{4\pi}\int d^2r\int_0^1\frac{z}{4\pi}\int d^2b\int d^2r^\prime
				\int_0^1\frac{z^\prime}{4\pi}\int d^2b^\prime(\Psi_V^*\Psi_\gamma)_{T}(z,r)(\Psi_V^*\Psi_\gamma)_{T}(z^\prime ,r^\prime )\notag\\&&\times e^{-i(b-b^\prime )\cdot \Delta} \Big\{1-S_{x_A}(x_1,x_2,b)-S_{x_A}(x^\prime_2,x^\prime_1,b^\prime)+S_{x_A}(x_1,x_2,b)S_{x_A}(x^\prime_2,x^\prime_1,b^\prime)\nonumber\\ &&+S_{x_A}(x_1,x_2,b)S_{x_A}(x^\prime_2,x^\prime_1,b^\prime)\frac{16(r\cdot r^\prime )^2}{N_c^2((r+r^\prime )^2-4(b-b^\prime )^2)}\nonumber\\ &&\times [e^{\frac{Q_S^2((r+r^\prime )^2-4(b-b^\prime )^2)}{8}}-\frac{Q_S^2((r+r^\prime )^2-4(b-b^\prime)^2)}{8}-1]\Big\}.
\label{nlocross}
	\end{eqnarray}
	We can see that the first two line of Eq.~(\ref{nlocross})  is the just the coherent cross section. The rest is the large-$N_c$ contributions. We think the large-$N_c$ contribution should contribute incoherent cross section in the high energy limits. The large-$N_c$ differential cross section reads 
	\begin{eqnarray}
	\frac{d\sigma^{\gamma A\to VX}}{dt}=&&\frac{R_g^2(1+\beta^2)}{4\pi}\int d^2r\int_0^1\frac{z}{4\pi}\int d^2b\int d^2r^\prime
			\int_0^1\frac{z^\prime}{4\pi}\int d^2b^\prime(\Psi_V^*\Psi_\gamma)_{T}(z,r)(\Psi_V^*\Psi_\gamma)_{T}(z^\prime ,r^\prime )\notag\\ &&\times  e^{-i(b-b^\prime )\cdot \Delta} S_{x_A}(x_1,x_2,b)S_{x_A}(x^\prime_2,x^\prime_1,b^\prime) \frac{16(r\cdot r^\prime )^2}{N_c^2((r+r^\prime )^2-4(b-b^\prime )^2)}\notag\\ &&\times \big[e^{\frac{Q_S^2((r+r^\prime )^2-4(b-b^\prime )^2)}{8}}-\frac{Q_S^2((r+r^\prime )^2-4(b-b^\prime)^2)}{8}-1\big].
	\label{in_ds}
	\end{eqnarray} 
	Using  Eqs. (\ref{in_ds}) and (\ref{sbs}),  we get the incoherent cross section. Finally, 
	 with  Eq.~(\ref{dndk}), we get the $J/\psi$ rapidity distributions of coherent and incoherent production  in PbPb ultraperipheral collisions. 	
\section{Numerical results}
\begin{figure}[h]
 	\begin{center}
 	\includegraphics[width=4in]{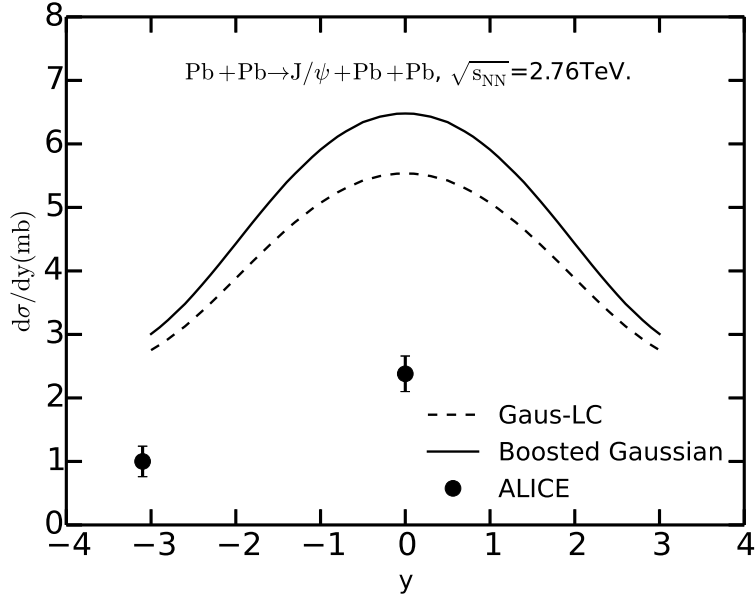}\hspace{3.0cm}
 	\caption{ The compared rapidity distributions of $J/\psi$ coherent production in PbPb collisions at $\sqrt{s_{NN}}=2.76$~TeV,  the circle are the experimental data of the ALICE collaboration~\cite{Abelev:2012ba,Abbas:2013oua}, the solid curve is used the Boosted Gaussian model, and dashed curve is used the Gaus-LC model.}\label{fig7}
 	\end{center}
 	\end{figure}

\begin{figure}[h]
 	\begin{center}
 	\includegraphics[width=4in]{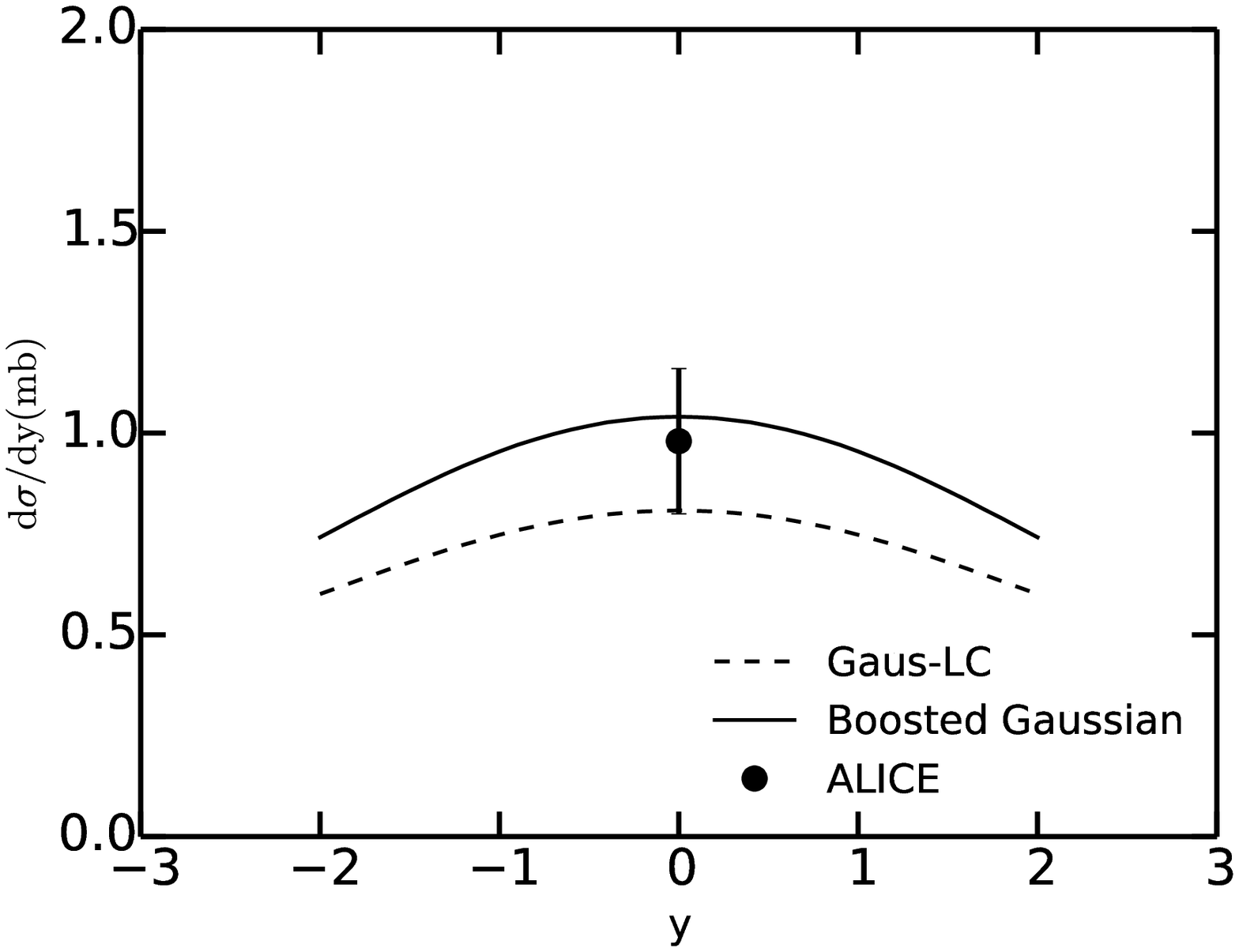}\hspace{3.0cm}
 	\caption{ The compared rapidity distributions of $J/\psi$ incoherent production in PbPb collisions at $\sqrt{s_{NN}}=2.76$~TeV,  the circle is the experimental data of the ALICE collaboration~\cite{Abelev:2012ba,Abbas:2013oua}, the solid curve is used the Boosted Gaussian model, and dashed curve is used the Gaus-LC model.}\label{fig8}
 	\end{center}
 	\end{figure}

The ALICE  collaboration had measured the coherent and incoherent production in PbPb ultraperipheral collision at $\sqrt{s_{NN}}=2.76$~TeV. The coherent and incoherent $J/\psi$ production in the $-0.9<y<0.9$ are $d\sigma^{coh}/dy=2.38^{+0.34}_{-0.24}~\mathrm{mb}$ and $d\sigma^{incoh}/dy=0.98^{+0.19}_{-0.17}~\mathrm{mb}$~\cite{Abbas:2013oua}. In the rapidity region $-3.6<y<2.6$, the coherent production is $d\sigma^{coh}/dy=1.00\pm0.18^{+0.24}_{-0.26}~\mathrm{mb}$~\cite{Abelev:2012ba}. The ratio of the incoherent to coherent is about $41.2^{+3.0}_{-2.1}\%$ at midrapidity.
In the Ref.~\cite{Lappi:2013am}, the authors had calculate the production of the coherent and the incoherent in PbPb ultraperipheral collision. The prediction of Ref.~\cite{Lappi:2013am} of coherent production is upper than the measurement of ALICE. The ratio of incoherent to coherent is $23\%$ using IPsat and Boosted Gaussian model, which is lower than the ratio of ALICE at midrapidty. 
We calculate the rapidity distributions of  $J/\psi$  production in PbPb ultraperipheral collisions  at $\sqrt{s_{NN}}~=~2.76~\mathrm{TeV}$, the following parameters are evolved in the calculations,  the lorentz boost factor is  $\gamma_L=\sqrt{s_{NN}}/2m_p=1482$,  the radius of Pb nucleus is $R_A=1.2A^{1/3}=7~\mathrm{fm}$, with $ \mathrm{A}=208$. The $Q_{S0}^2=1.0~\mathrm{GeV}^2$, $x_0=3.04\times 10^{-4}$,  $\lambda=0.229$, $c(b) = 0.312$. The mass of $J/\psi$ is $M_V=3.097$~GeV,  the mass of the charm quark is $m_f=1.4$~GeV,  and $Q^2=0$~GeV  for the overlap of wave functions. \\
The parameters of GBW model are not very reliable because they are fit from the inclusive production of $\gamma^*+p\to \gamma^*+p$. As there are no experimental data in the $\gamma^*+A\to \gamma^*+A$. In the following calculation, the ratio between the incoherent and coherent is reliable.\\
\indent We present the theoretical results and experimental data of $J/\psi$ coherent and incoherent rapidity distributions in PbPb ultraperipheral collisions at $ \sqrt{s_{NN}}=2.76~\mathrm{TeV}$  in  Fig.~\ref{fig7}and Fig.~\ref{fig8}. \\
\indent The prediction of coherent production of $J/\psi$  are shown in Fig.~\ref{fig7}. The circle are experimental data measured by the ALICE collaboration~\cite{Abelev:2012ba,Abbas:2013oua}. The solid curve  and the  dashed curve are predictions  calculated in Boosted Gaussian and Gaus-LC model.  The range of integrated transfer momentum is $0<|t|<0.1~\mathrm{GeV}^2$.  We can see the prediction of ours is larger than the experimental data. Because the parameters of our model is not reliable. \\
 \indent Now, let's do the calculation of incoherent production from Eq.~(\ref{in_ds}). In the calculation, we define $\Delta b=b-b^\prime$, and we suppose that $\Delta b\ll b$. Then, we can take the approximation $S(x_1^\prime,x_2^\prime,b^\prime)\approx S(x_1^\prime,x_2^\prime,b)$ in the calculation of incoherent prodcution. The prediction of incoherent production of $J/\psi$  are shown  Fig.~\ref{fig8}. The circle is the experimental data  measured by the ALICE collaboration~\cite{Abbas:2013oua}. The range of integrated momentum is $\mathrm{0.1~GeV}^2<|t|<\mathrm{0.3~GeV}^2$. We can see that the prediction of the incoherent give a good description to the data of ALICE. But the coherent prediction is not good description to the data of ALICE.\\
 \indent At the end of the day, let's consider the ratio of incoherent to the coherent production, because the ratio of incoherent to the coherent is reliable. In our calculation, the ratio of incoherent to coherent prediction at midrapidity is $15\% $ using Gaus-LC model and $16\%$ using Boosted Gaussian at midrapidity. The results of ratio of incoherent to the coherent are presented in Table.~\ref{IPP}, where we add the ratio of Ref.~\cite{Lappi:2013am}. We can see that the sum of the IPsat in Ref.~\cite{Lappi:2013am} and this work give a good description of the ratio to the experimental ratio.
  \begin{table}[h]
  \begin{center}
  \begin{tabular}{|c|c|c|c|c|c|}
  \hline
  \hline
  & IPsat model& IIM model  &This work& IPsat+This work & IIM+This work \\
  \hline
 BG &$22\%$ &  $14\%$&$16\%$  & $38\%$ &  $30\%$    \\
 \hline
Gaus-LC  & $21\%$ &$13\%$& $15\%$ &  $36\%$		    & $28 \%$      \\
   \hline
ALICE &  & 	&    &$41.2^{+3.0}_{-2.1}\%$ &   $41.2^{+3.0}_{-2.1}\%$  \\
  \hline\hline
  \end{tabular}
  \caption{The prediction of ratio of incoherent to coherent at midrapidty at PbPb ultraperipheral collsion, the prediction of IPsat and IIM model are taken from Ref.~\cite{Lappi:2013am}. BG is the Boosted Gaussian wave function.}
  \label{IPP}
\end{center}
 \end{table}
 \section{conclusions} 
 In summary,  in this manuscript, we calculate the photon flux in the light-cone perturbative theory, and we get the same result as the classic electrodynamics. We calculate the production of $J/\psi$ in PbPb ultraperipheral collision at $\sqrt{s_{NN}}=2.76$~TeV. The calculation is in the dipole picture. The Gaus-LC and Boosted Gaussian model are used in the forward wave function. and the GBW model is implemented in dipole cross section.
 We distinguish the coherent and incoherent cross section in the double gluon exchange and one gluon exchange. In the process where the nucleon exchange two gluon with the dipole, which is coherent process. In the process where the nucleon exchange one gluon with the dipole, we think it is incoherent process. In this work, the prediction of coherent is larger than the experimental data, but the ratio of the incoherent and coherent is reliable. We compute the ratio of the incoherent to the coherent, which is about $15 \%$ and $16 \%$. We think the prodction of this work and Ref.~\cite{Lappi:2013am} both contribute the incoherent production in PbPb ultraperipheral collision.  
\section{Acknowledgements}
One of the authors, Y. P. Xie, thanks Dr. Bo-Wen ~Xiao
for useful comments and discussions, and also thanks  communication with T.~Lappi and H.~Mantysaari. This work is supported
in part by the National Natural Science Foundation
of China (Grant No. 11175220), the One Hundred Person
Project (Grant No. Y101020BR0), and the Key Laboratory
of Quark and Lepton Physics (MOE), Central China
Normal University (Grant No. QLPL201414).

\end{document}